\documentclass[a4paper,10pt]{article}

\usepackage[pdftex]{graphicx}
\usepackage{amssymb,amsfonts,amsmath}
\usepackage{gensymb}
\usepackage{multicol}
\usepackage{cite}
\usepackage{textcomp}
\usepackage{upgreek}
\usepackage[latin1]{inputenc}
\usepackage{abstract}

\begin{document}

\author{Stefan Geissbuehler$^{1*}$,
Azat Sharipov$^1$,
Aurélien Godinat$^2$,
Noelia L. Bocchio$^1$,\\
Elena A. Dubikovskaya$^2$,
Theo Lasser$^1$, and Marcel Leutenegger$^1$}

\title{Multiplane 3D superresolution optical fluctuation imaging} 
\date{$^1$École Polytechnique Fédérale de Lausanne, Laboratoire d'Optique Biomédicale, Lausanne, Switzerland\\$^2$École Polytechnique Fédérale de Lausanne, Chaire Neva de chimie bio-organique et d'imagerie moléculaire, Lausanne, Switzerland}
\maketitle
\vspace{0.3cm}

\begin{abstract}
By switching fluorophores on and off in either a deterministic \cite{Hell:1994tt,Gustafsson:2005ts} or a stochastic \cite{Rust:2006wn,Heilemann:2008vd,Betzig:2006hx,Hess:2006wl,Dertinger:2009wh} manner, superresolution microscopy has enabled the imaging of biological structures at resolutions well beyond the diffraction limit. Superresolution optical fluctuation imaging (SOFI) \cite{Dertinger:2009wh,Dertinger:2010wt} provides an elegant way of overcoming the diffraction limit in all three spatial dimensions by computing higher-order cumulants of image sequences of blinking fluorophores acquired with a conventional widefield microscope. So far, three-dimensional (3D) SOFI has only been demonstrated by sequential imaging of multiple depth positions \cite{Dedecker:2012tj,Dertinger:2012ce}. Here we introduce a versatile imaging scheme which allows for the simultaneous acquisition of multiple focal planes. Using 3D cross-cumulants, we show that the depth sampling can be increased. Consequently, the simultaneous acquisition of multiple focal planes reduces the acquisition time and hence the photo-bleaching of fluorescent markers. We demonstrate multiplane 3D SOFI by imaging the mitochondria network in fixed C2C12 cells over a total volume of $65\times65\times3.5~\upmu\textrm{m}^3$ without depth scanning.
\end{abstract}

\begin{multicols}{2}
\paragraph{}
The optical performance of a microscope is characterized by its 3D pointspread function (PSF) whose spatial extents determine the resolution in the corresponding spatial dimensions. The image of a single fluorescent molecule essentially corresponds to the PSF centered on the position of the molecule. By sequentially imaging isolated fluorophores, determining their PSF centers and mapping them into a position histogram, localization microscopy can render the sample structure with much finer details than the classical fluorescence microscope \cite{Rust:2006wn,Heilemann:2008vd,Betzig:2006hx,Hess:2006wl}. However, this requires the use of fluorophores that can be photo-switched or -activated stochastically. The characteristic rates of activation and deactivation need to be tunable such that only a sparse subset of isolated molecules is active per exposure and a sufficiently high photon number can be collected for a precise localization \cite{vandeLinde:2010kg}. As an alternative to single-molecule localization, SOFI achieves superresolution using higher-order statistics with relaxed requirements on the photo-switching kinetics \cite{Dertinger:2009wh,Geissbuehler:2011vh}. Unlike localization microscopy, SOFI does not require the spatio-temporal isolation of individual fluorophore emission patterns nor a complete on-off switching. SOFI only relies on reversible stochastic and independent intensity fluctuations from marker molecules. It has been successfully demonstrated with semiconductor quantum dots \cite{Dertinger:2009wh,Dertinger:2010wt}, organic dyes \cite{Dertinger:2010ta} and fluorescent proteins \cite{Dedecker:2012tj}. With the help of time-correlated single photon counting instrumentation, the stochastic photon emission from single fluorophores and the associated antibunching effect was exploited in combination with a modified SOFI analysis \cite{Schwartz:2012ts}. Recently, SOFI was applied to fluorescence fluctuations induced by stochastic approaches of fluorescence resonance energy transfer (FRET) pairs \cite{Cho:2013fl}, where the FRET donor was linked to the structure of interest and the acceptor diffusing in the medium.

\begin{figure*}[ht]
\centerline{\includegraphics[width=8.8cm]{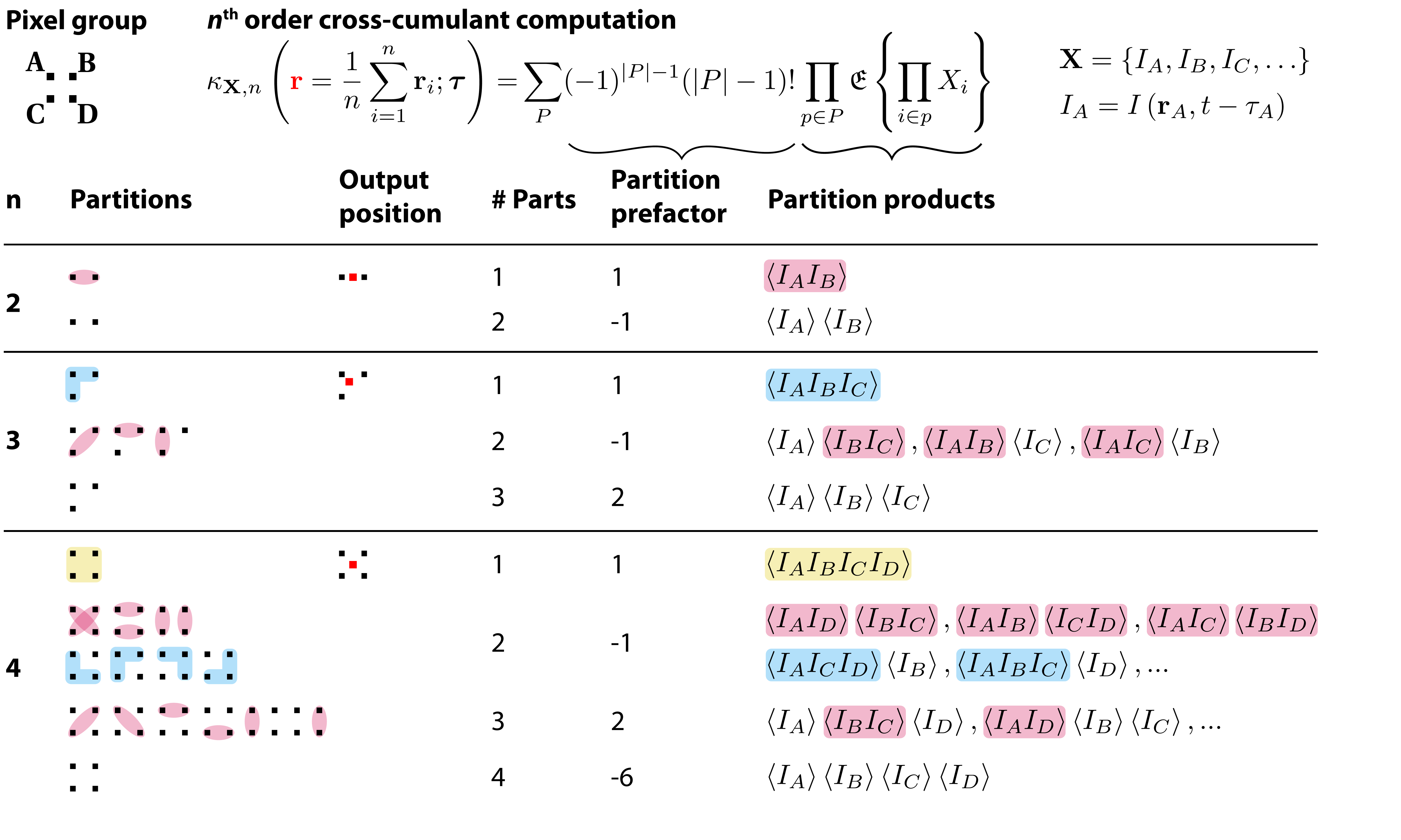}}
\caption[Calculation of cross-cumulants]{Calculation of cross-cumulants. The cross-cumulant $\kappa_{\mathbf{X},n}$ of order $n$ for a pixel group $\mathbb{S}=\left\{\mathbf{r}_A,\mathbf{r}_B,\mathbf{r}_C,\mathbf{r}_D\right\}$ can be calculated as a sum over all partitions of $\mathbb{S}$. Each addend consists of a prefactor depending on the number of parts in the particular partition times a partition product which evaluates the parts as a temporally averaged $\left<...\right>$ product.}
\label{crosscumulantscalc}
\end{figure*}

The simplest implementation of SOFI corresponds to the calculation of the temporal variance of an image sequence of stochastically blinking emitters. Because the blinking between different emitters is uncorrelated, the variance of the sequence from a sum of $N$ emitters is the sum of the variances of every single-emitter sequence $I_i$:

\begin{equation}
\textrm{var}\left\{\sum_{i=1}^NI_i(\mathbf{r},t)\right\}=\sum_{i=1}^N\textrm{var}\left\{I_i(\mathbf{r},t)\right\},
\end{equation}
where $i$ is the index of the emitter, $\mathbf{r}$ denotes a pixel location on the detector and $t$ is the time. The signal from a single emitter located at $\mathbf{r}_i$ corresponds to a sampled and shifted version of the microscope's PSF $U\left(\mathbf{r}-\mathbf{r}_i\right)$ with brightness amplitude $\epsilon_i$ times a fluctuation signal $s_i(t)$ that encodes the temporal fluctuations. The single-emitter variance is

\begin{equation}
\textrm{var}\left\{I_i(\mathbf{r},t)\right\}=\epsilon_i^2U^2\left(\mathbf{r}-\mathbf{r}_i\right)\textrm{var}\left\{s_i(t)\right\}.
\end{equation}
Hence, the variance-equivalent PSF is given by the original PSF raised to the power of two. This doubles the spatial frequency support of the corresponding optical transfer function (OTF) and therefore improves the spatial resolution.

Instead of calculating the temporal variance on each pixel separately, the covariance of neighboring pixels proves to be beneficial as it cancels any uncorrelated noise. Furthermore, by using different combinations of pixel pairs for calculating the covariance, the final image sampling can be increased as the covariance essentially probes the sample in the geometrical mean of the chosen pixel pair. 

The concept of multiplying the PSF with itself for improving the resolution as in the variance and covariance can be generalized to higher orders using cumulants or cross-cumulants. The second-order cumulant is equivalent to the variance and the second-order cross-cumulant is identical to the covariance. An $n$-th order cumulant or cross-cumulant raises the PSF to the power of $n$. Cumulants thus provide a resolution that is ultimately limited by the signal-to-noise ratio only. Cross-cumulants among a group of pixels allow to eliminate noise and to increase the pixel grid density, such that the cumulant image resolution is not limited by the effective pixel size \cite{Dertinger:2010wt}. This results in a 3D image with up to $n^3$ times the original number of pixels for a cumulant order $n$. Figure 1 illustrates the calculation of cross-cumulants up to the fourth order for pixels $\left\{\mathbf{r}_A,\mathbf{r}_B,\mathbf{r}_C,\mathbf{r}_D\right\}$ that may differ in any spatial direction including the so far neglected axial dimension. To maximize the speed and the field-of-view, each dimension of the PSF is oversampled to satisfy Shannon's sampling theorem \cite{Shannon:1998uo} while ensuring a correlated signal between the cross-cumulated pixels. In general, the more pixels that are acquired simultaneously, the faster the overall image acquisition becomes. It is thus advantageous to use a widefield imaging system with a high pixel-count camera that is sufficiently sensitive and rapid to detect the emitters' fluorescence fluctuations.

In order to fully exploit the resolving power of higher-order cross-cumulants with a balanced image contrast, we applied a simplified linearization \cite{Geissbuehler:2012dr}. The absolute value of the resulting cumulant image is first deconvolved in 3D using a Lucy-Richardson algorithm. The log-likelihood formulation used in the algorithm is well adapted for cumulants as long as there are no sign changes within a PSF. We then take the $n$-th root and reconvolve the image with an $n$-times size-reduced PSF to limit the apparent final resolution to a physically reasonable value given by the support of the cumulant OTF.

\begin{figure*}[ht]
\centerline{\includegraphics[width=13cm]{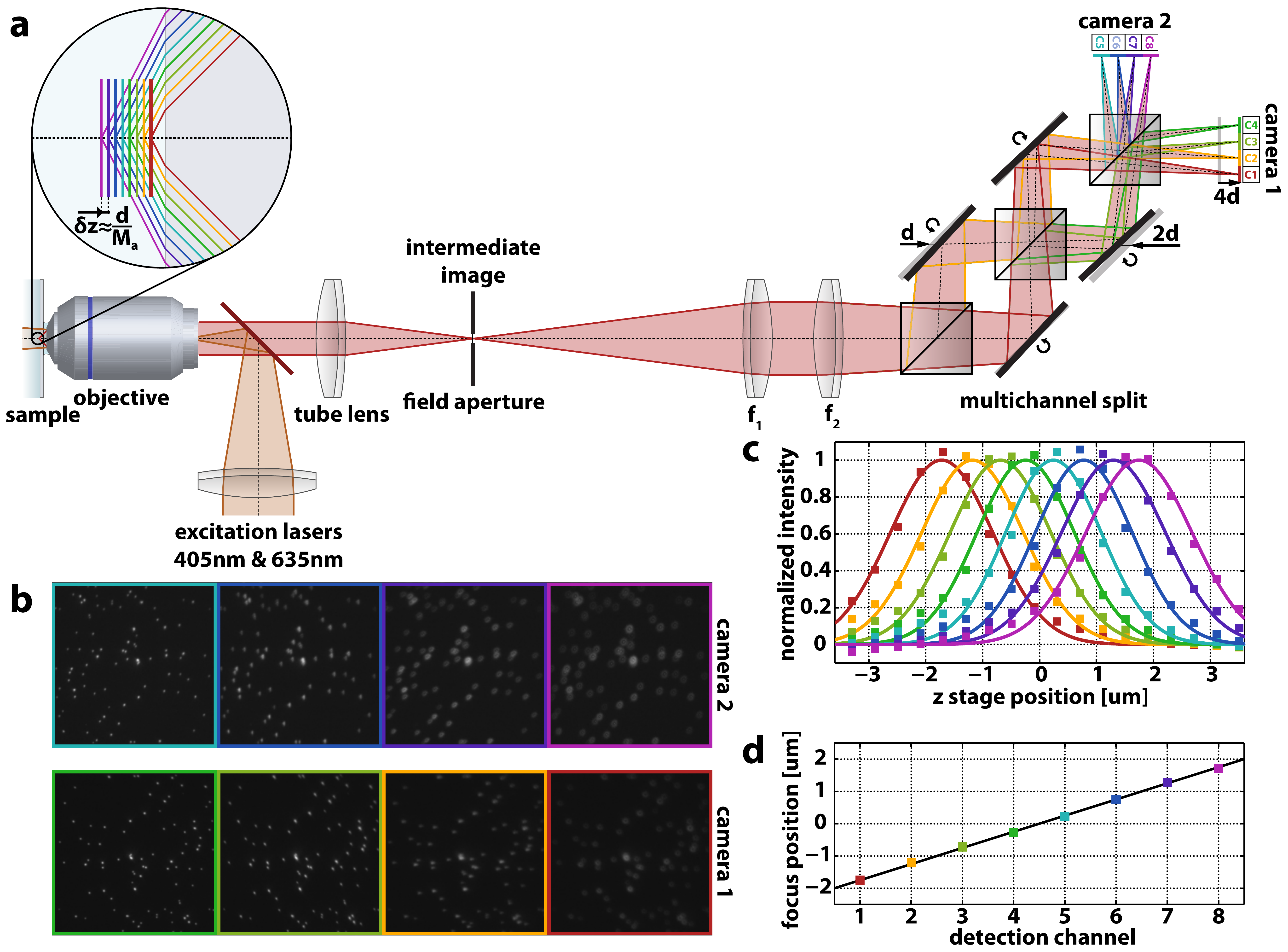}}
\caption[Detection scheme for the simultaneous acquisition of multiple focal planes and/or colors]{Simultaneous detection of multiple focal planes. (a) Multiplane widefield fluorescence microscope. The excitation laser is focused into the aperture of the objective to create a large illumination field. Sample features in several focal planes are imaged simultaneously by two cameras. These focal planes are obtained by splitting the fluorescence into several channels using three 50:50 beam splitter cubes and by introducing path length differences $d$ and $2d$ with the mirrors, as well as $4d$ with camera 1. The resulting separation between the focal planes in the object space is then approximately $d/M_a$, with $M_a$ being the axial magnification of the system. The images of the focal planes are projected side-by-side on the cameras. A field stop in the intermediate image prevents overlaps between the image frames. (b) Image frames taken with a fluorescent bead sample. (c) Peak brightness in the image frames in (b) upon scanning the sample plane through the focal channels. (d) Calibration of the focal plane separation by the positions of the brightness maxima in (c).}
\label{multiplane_concept}
\end{figure*}

Figure 2.a shows the schematic of a widefield setup for the simultaneous acquisition of multiple focal planes with two cameras. The use of an index-matched immersion objective lens (e.g. a water immersion objective if the sample is in an aqueous environment) minimizes spherical aberrations when imaging thick samples. Three diagonally arranged 50:50 beamsplitters behind the lens $f_2$ split the collected fluorescence into eight separate detection channels. Four mirrors are oriented in such a way as to direct the light onto separate areas of the two cameras. By axially shifting two mirrors by a distance $d$ and $2d$ as indicated in Fig. 2.a and by shifting one camera by a distance $4d$, eight different optical path lengths are achieved with respect to the lens $f_2$. Therefore, eight distinct and conjugated depth positions are probed. The focal plane separation in the object space is then approximately given by $d/M_a$, where $M_a$ is the axial magnification. To comply with Shannon's sampling theorem, this distance should not exceed half the axial resolution of the imaging system. The field stop has an adjustable rectangular shape and is used to avoid cross talk between the channels. By adding or removing a pair of mirrors and a beamsplitter, the number of detection channels can be increased or decreased, respectively.

\paragraph{}
We calibrated the distances between the focal planes in the sample to $500~\textrm{nm}$ (Fig. 2.c and d) and determined the transverse positioning and the mutual distortion of the detection channels for coregistration. Using a fluorescent bead sample (Fig. 2.b) we determined the full-width-at-half maxima of the PSF in the transverse ($320~\textrm{nm}$) and axial dimension ($1.2~\upmu\textrm{m}$).

\begin{figure*}[ht]
\centerline{\includegraphics[width=14cm]{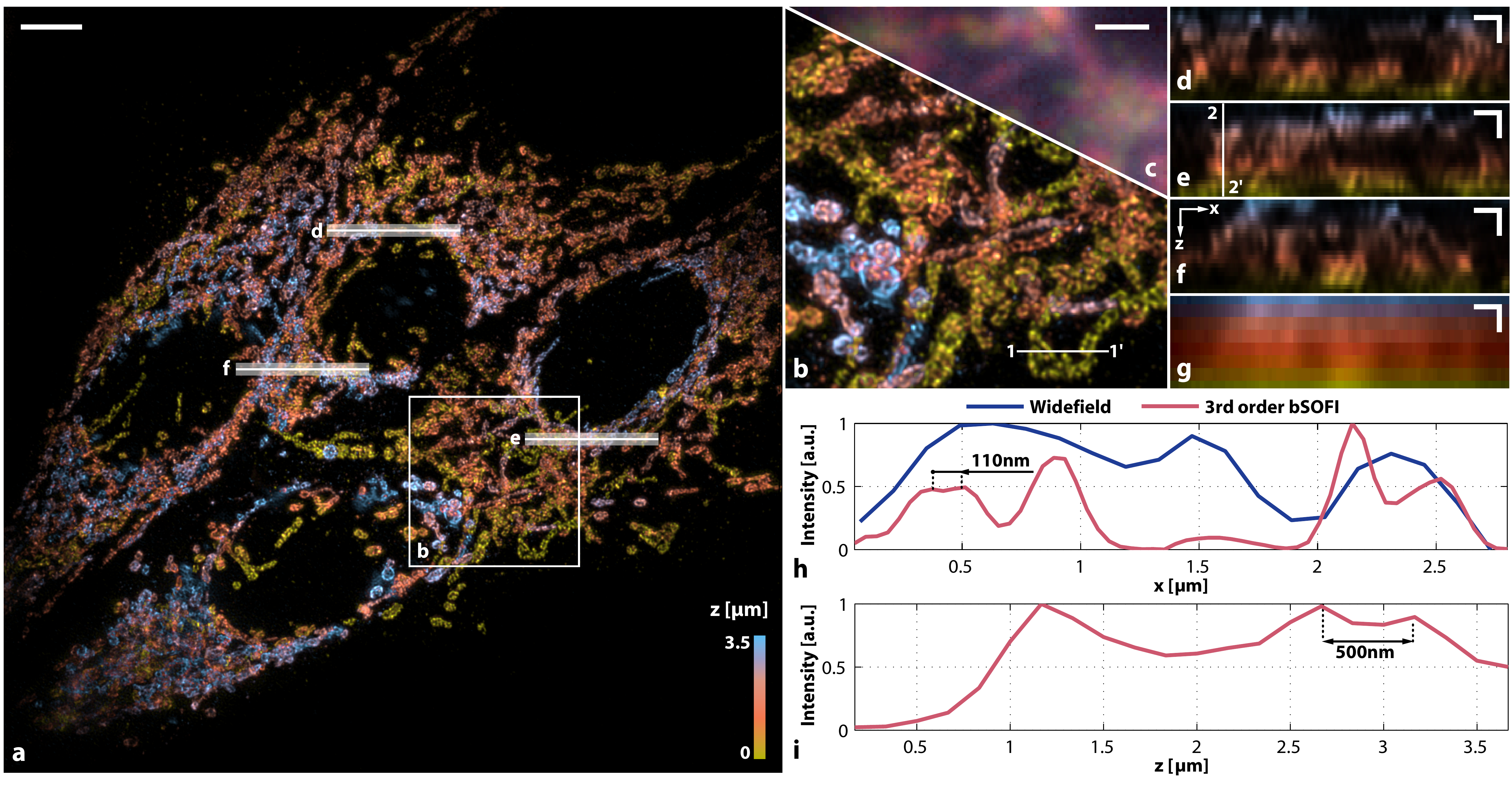}}
\caption[Mitochondria in C2C12 cells]{Mitochondria stained with Alexa 647 in fixed C2C12 cells. (a) Maximum intensity projection of the third-order balanced SOFI image covering a volume of $65\times65\times3.5~\upmu\textrm{m}^3$. (b) Zoomed region highlighted in (a). (c) Diffraction-limited widefield view of the same area as (b). (d-f) x-z sections highlighted in (a) avaraged over 25 pixels along y. (g) Same x-z section as (f) from diffraction-limited widefield stack. (h) x-profiles from the cut 1-1' highlighted in (b). (i) z-profile from the cut 2-2' highlighted in (e). {\it Scale bars: (a) $5~\mu\textrm{m}$ (b,c) $2~\mu\textrm{m}$ (d-g) $1~\mu\textrm{m}$ in x and z. For the display of the z coordinate we used an ``isolum''  colormap compatible with red-green color perception deficiencies \cite{Geissbuehler:2013er}.}}
\label{fig:c2c12_mitochondria_mpsofi.pdf}
\end{figure*}

We imaged the outer membrane of mitochondria in fixed C2C12 cells using antibody staining with Alexa 647. Taking 5000 images acquired at 40 frames per second, we obtained 3D superresolved images of mitochondria as shown in Fig. 3 and Supplementary Video 1. The simultaneous acquisition of multiple focal planes allowed a volume of $65\times65\times3.5~\upmu\textrm{m}^3$ to be imaged without scanning. We computed third-order cross-cumulants in order to increase the sampling density three-fold in all spatial dimensions. Instead of the original 8 sampling points along the optical axis (Fig. 3.g), we therefore obtained 22 points, which significantly improves the visibility of structural details in depth (Fig. 3.d-f). The third-order balanced cumulant image \cite{Geissbuehler:2012dr} clearly reveals spherical and tubular mitochondrial features (Fig. 3.a,b,d-f), whereas in the diffraction-limited image those features are barely visible (Fig. 3.c,g). The strong out-of-focus background is almost completely removed in the 3D SOFI image. As shown in the profiles in Fig. 3.h and 3.i, the smallest features resolved in transverse and meridional sections of the image suggest a lateral resolution of $\sim110~\textrm{nm}$ and an axial resolution better than $500~\textrm{nm}$, which comes close to the expected three-fold increase in 3D resolution.

\paragraph{}
For widefield 3D SOFI, instead of acquiring the axial sections sequentially, it is favorable to acquire multiple depth planes simultaneously. Using a cross-cumulants analysis of stochastically blinking fluorescent emitters across the individual depth planes, the physically acquired depth planes are supplemented by virtual planes that provide a finer depth sampling. As a consequence, the axial PSF does not have to be $n$ times oversampled as would be the case for axially scanned $n$-th order 3D SOFI. Accordingly, multiple simultaneously acquired focal planes in combination with 3D cross-cumulants allow for a reduction in the acquisition time and a better exploitation of the available photon budget from the fluorescent markers, particularly if the emitters bleach during the acquisition.

As a further consequence, the resulting contrast of the 3D cross-cumulants is expected to be more homogeneous along the optical axis when compared to sequential plane-by-plane 2D cross-cumulants. During sequentially acquired depth sections, the cumulant response of individual emitters might change due to stochasticity resulting in a potentially imbalanced cumulant contrast which would be problematic for subsequent 3D deconvolution and cumulant balancing (as described in \cite{Geissbuehler:2012dr}).

Multiplane 3D SOFI is a fast and versatile widefield imaging concept allowing superresolved cell imaging in 3D without scanning. We imaged the outer membrane of mitochondria in C2C12 cells with an eight-plane detection scheme covering $3.5~\upmu\textrm{m}$ in depth and demonstrated an almost three-fold resolution improvement in all three dimensions as compared to the diffraction-limited image.

\section*{Materials}
\subsection*{Microscope setup} 
The custom-built microscope system (Figure 2.a) comprised a 1.2NA 60$\times$ water immersion objective (UPLSAPO 60XW, Olympus), two diode lasers for excitation and reactivation (MLL-III-635 800~mW, Roithner Lasertechnik and iBeam smart 405 120~mW, Toptica) adjusted to illumination intensities of 2 kW/cm$^2$ at 635~nm and 20~W/cm$^2$ at 405~nm, a multiplane splitter with three 50:50 beamsplitters (BS013, Thorlabs) and two sCMOS cameras (ORCA Flash 4.0, Hamamatsu).

\subsection*{Data analysis} A detailed description of the multiplane SOFI analysis may be found in Supplementary Methods with a flowchart given in Supplementary Figure 1. In brief, each registered pair of image sequences from consecutive depth channels was divided into 500-frame sub-sequences and processed with a 3D cross-cumulants analysis using voxel combinations without repetitions (Supplementary Figure 2) and zero time delays. The frames of one depth sequence were analyzed without preprocessing, whereas the frames of the other sequence were coregistered using an affine transformation and bilinear interpolation. The transformation parameters were obtained from a calibration measurement using fluorescent beads. A subsequent fine-tuning of the image registration using a rigid transformation on the measurement data was necessary to correct for residual mechanical drift of the detection optics. From each bi-plane cross-cumulants analysis, only virtual planes and the untransformed physical plane were kept for further processing, because the transformed depth plane contains correlated noise parts in neighboring pixels introduced by the interpolation. The different detection responses in the channels were adapted by using a ratio histogram analysis of the two virtual planes obtained by a third order analysis. A distance factor that maximized the smoothness of the images was applied to correct for the different pixel weights in the raw cumulant images. Subsequently, the bi-plane cumulant image blocks are transformed using the previously determined transformation masks in order to have all planes aligned. Finally, the nonlinear response of the cumulant with respect to differences in molecular brightness and blinking was linearized by using a 3D Lucy-Richardson deconvolution with 50 iterations, taking the $n$-th root and reconvolving with an $n$-times size-reduced PSF.

\subsection*{Cell culture} C2C12 cells were cultured at 37\degree{C} with 5\% CO$_2$ using Dulbecco's Modified Eagle Medium (DMEM) high glucose (4.5~g/l of glucose), containing 10\% FBS, 2\% HEPES 1~M, 1\% MEM Non-Essential Amino Acids Solution (100$\times$) and 40~$\upmu$g/ml of Gentamycin (all products were purchased from Life Technologies Corporation, Paisley, UK). Cells were plated in 35~mm Fluorodish Cell Culture Dish (World Precision Instruments Inc., Sarasota, US) at roughly 50000 cells per plate on the day before fixation.

\subsection*{Immunostaining} C2C12 cells were washed 3 times carefully with PBS and were then incubated for 10 minutes at room temperature in 10\% neutral buffered Formalin solution (4\% w/v formaldehyde, Sigma-Aldrich, St. Louis, US) for fixation, followed by washing 3 times 5 minutes with PBS. The cells were then incubated for 60 minutes in TBS containg 0.1\% Tween-20 (TBST) with 10\% fetal calf serum (FCS)  at room temperature, followed by incubation overnight at 4\degree{C} with a primary antibody for TOMM20 (mouse monoclonal, ABCAM ab56783) diluted 1:100 in TBST with 10\% FCS. Cells were then washed 3 times 5 minutes with PBS and incubated again at room temperature. for 1-2h with a secondary antibody (donkey anti-mouse, Alexa Fluo 647), diluted 1:150 in TBST with 10\% FCS. After washing the cells 3 times for 5 minutes with PBS, a postfixation was performed for 10 minutes with Formalin solution at room temperature. Before addition of a 1:1 PBS/Glycerol solution, the cells were finally washed again 3 times for 5 minutes with PBS.

\subsection*{Imaging buffer} The samples were imaged using a buffer containing an oxygen scavenger and a thiol. The solutions were prepared using phosphate-buffered saline (PBS) (Sigma Aldrich),  0.5~mg/ml glucose oxidase (Sigma Aldrich), 40~$\upmu$g/ml catalase (Sigma Aldrich), 6.6\%~w/v glucose and 10~mM cysteamine (Sigma Aldrich).

\section*{Acknowledgments}
The authors would like to thank Prof. Jo\-han Au\-werx for kindly providing the C2C12 cells, Dr. Erica Martin-Williams for proof-reading the manuscript, Prof. Gabriel Popescu and Dr. Claudio Dellagiacoma for helpful comments and discussions and Gennady Nikitin for his assistance with cell labeling. This research was supported by the Swiss National Science Foundation (SNSF) under grants CRSII3-125463/1 and 205321-138305/1. A.G. and E.A.D. acknowledge the grants from Neva Foundation.

\section*{Author contributions}
S.G., E.A.D., T.L. and M.L. conceived the study. S.G. and M.L. developped the algorithms and A.G., N.L.B. and S.G. prepared the samples. S.G. and A.S. performed experiments and analyzed data. S.G. wrote the manuscript and all authors commented on it at every stage.

\section*{Competing financial interests}
The authors declare no competing financial interests.

\end{multicols}
\end{document}